\def\Roman#1{\uppercase\expandafter{\romannumeral#1}}
\documentclass[a4paper,12pt]{article}
\usepackage{amssymb,amsfonts}
\usepackage[mathscr]{eucal}

\title{On  geometrical representation of the Jacobian in a path integral reduction problem}
\author{S. N. Storchak\\
\small{Institute for High Energy Physics, Protvino, Moscow Region,142284,Russia}}

\begin{document}

\maketitle

\begin{abstract}
The geometrical representation of the Jacobian in the path integral reduction problem which describes a motion of the scalar particle on a smooth compact Riemannian  manifold with the given free isometric action of the compact semisimple Lie group is obtained. By using the formula for the scalar curvature of the manifold with the Kaluza--Klein metric, we   present the Jacobian   as difference of the scalar curvature of the total space of the principal fibre bundle and the terms that are
the scalar curvature of the orbit space, the scalar curvature of the orbit, the second fundamental form of the orbit and the square of the principle fibre bundle curvature.  
\end{abstract}

\section{Introduction}
In our papers \cite{Storchak_1,Storchak_2} we have developed an approach to the factorization of the path integral measure in  Wiener path integrals that can be  used in  
the Euclidean quantization of the finite--dimensional dynamical systems with a symmetry. 
With the path integrals of these papers we 
 represented the solutions of the  backward Kolmogorov equations that correspond (via changing an appropriate real parameter of the equation for the complex i) to the Schr\"odinger equations.

In the dynamical system which describes the  motion of a scalar particle on the compact Riemannian manifold with the given free isometric action of a compact semisimple Lie group we studied the path integral reduction problem. That is, we considered the transformation of the original path integral which leads  to the path integral for a new dynamical system given  on the reduced space.

In our papers, the path integrals  on the manifold and on the principal fibre bundle were defined by the Belopolskaya and Daletskii method \cite{Daletskii}. By this method, the integration measures of the path integrals are generated by the stochastic processes that are given on a manifold. 
The stochastic processes are determined by the solutions of the stochastic differential equations. Considering these equations (and their solutions) on charts of the manifold, it is possible to define the local evolution semigroups 
acting in the space of functions given on the manifold. 

Every local semigroup can be represented as a path integral whose  path integral measure is defined by the probability distribution of the local stochastic process. This process is a local representative of the global stochastic process.
The limit of the superposition of these semigroups leads to the global semigroup which determines the global path integral.

 The solution of  the backward Kolmogorov equation on a smooth compact
Riemannian manifold $\mathcal P$:
\begin{equation}
\left\{
\begin{array}{l}
\displaystyle
\left(
\frac \partial {\partial t_a}+\frac 12\mu ^2\kappa \,
\triangle _{\mathrm P}
(Q_a)+\frac
1{\mu ^2\kappa \, m}V(Q_a)\right)\psi (Q_a,t_a)=0\\
\psi (Q_b,t_b)=\varphi _0(Q_b)
\qquad\qquad\qquad\qquad\qquad (t_{b}>t_{a}),
\end{array}\right.
\label{1}
\end{equation}
in which $\triangle _{\mathrm P}(Q)=G^{-1/2}\frac \partial {\partial Q^A}G^{AB}G^{1/2}\frac
\partial {\partial Q^B}
$\footnote{The indices denoted by capital letters  run from 1 to $n_{\mathrm P}
=\dim {\mathcal P}$}
is the Laplace--Beltrami operator on 
$\mathcal P$, 
 $G=\det G_{\mathrm {AB}}$,
$\mu ^2=\frac \hbar m$ and $\kappa $ is a real positive parameter,
  can be presented as follows:
\begin{eqnarray}
\psi (Q_a,t_a)&=&{\rm E}\left[\varphi _0(\eta (t_b))\exp \left\{\frac 1{\mu
^2\kappa\, m}\int_{t_a}^{t_b}V(\eta (u))du\right\}\right]\nonumber\\
&=&\int_{\Omega _{-}}d\mu ^\eta (\omega )\varphi _0(\eta (t_b))\exp \{\ldots 
\},
\label{2}
\end{eqnarray}
where the path integral measure $\mu ^\eta $ is defined on the path space 
$\Omega _{-}=\{\omega (t):\omega (t_a)=0,\eta (t)=Q_a+\omega (t)\}$ 
given on the manifold $\mathcal P$.

The local representative ${\eta}_t^A$
of the global  process ${\eta}_t$ are determined by the solution of the following stochastic differential equation
\begin{equation}
d\eta ^A(t)=\frac 12\mu ^2\kappa G^{-1/2}\frac \partial {\partial
Q^B}(G^{1/2}G^{AB})dt+\mu \sqrt{\kappa }{\mathcal X}_{\bar{\mathrm M}}^A(\eta (t))dw^{
\bar{M}}(t)\\
\label{3}
\end{equation}
(${\mathcal X}_{\bar{\mathrm M}}^{\mathrm A}$ is defined by a local equality 
$\sum^{n_{\mathrm P}}_{{\bar{\mathrm K}}=1}{\mathcal X}_{\bar{\mathrm K}}^{\mathrm A}
{\mathcal X}_{\bar{\mathrm K}}^{\mathrm B}=G^{AB}$,
 and  we denote the Euclidean indices by over-barred indices).

The original manifold $\cal P$ of our dynamical system can be viewed locally  as a fibre space and we come to the principal fibre bundle $\pi :\mathcal P\rightarrow 
{\mathcal P}/{\mathcal G}=\mathcal M$, 
 where $\mathcal M$  is 
an orbit space of the right action of the group $\mathcal G$ on $\mathcal P$.
The total space of the bundle is our original manifold. Moreover, in this principal bundle there is a natural connection formed  by the   metric of the manifold.     

Using the transformation $Q^A=F^A(Q^{\ast}(x^i),a^{\alpha})$, 
we  change the  coordinates $Q^A$ of the manifold 
$\mathcal P$ for the adapted coordinates
$(x^i, a^{\alpha})$, where $x^i$ are the coordinates on the orbit space $\cal M$ and $a^{\alpha}$ -- the group coordinates of the fibre. As a result, we get the following representation of 
 the right invariant metric  $G_{AB}$: 
\begin{equation}
\displaystyle
\left(
\begin{array}{cc}
h_{ij}(x)+A_i^\mu (x)A_j^\nu (x){\gamma }_{\mu \nu }(x) & A_i^\mu
(x) 
\bar{u}_\sigma ^\nu (a){\gamma }_{\mu \nu }(x) \\ A_i^\mu (x) 
\bar{u}_\sigma ^\nu (a){\gamma }_{\mu \nu }(x) & \bar{u}%
_\rho ^\mu (a)\bar{u}_\sigma ^\nu (a){\gamma }_{\mu \nu }(x) 
\end{array}
\right)
\label{4}
\end{equation}
We see that  the original metric $G_{AB}$ becomes the Kaluza--Klein metric.

The orbit space metric $h_{ij}(x)$  of (\ref{4})
is defined by 
the formula:
\[
h_{ij}(x)={}^HG_{AB}(Q^{\ast}(x))\,\frac{\partial {Q^{\ast}}^A}{\partial x^i}\,
\frac{\partial {Q^{\ast}}^B}{\partial x^j},
\]
in which ${}^HG_{AB}=G_{CD}\,{\Pi}^C_A\, {\Pi}^D_B$. 
The projector 
${\Pi}^A_B={\delta}^A_B-K^{A}_{\alpha}d^{\alpha \beta}K_{\beta B}$
consists of the   Killing vectors 
$K^{A}_{\alpha}(Q)\frac{\partial}{\partial Q^A}$ and the metric 
along the orbits $d_{\alpha \beta}=K^{A}_{\alpha}G_{AB}K^{B}_{\beta}$. 
In order to define  ${}^HG_{AB}(Q^{\ast}(x))$ one must restrict 
the projectors ${\Pi}^C_A(Q)$ and the metric $G_{AB}(Q)$ to the orbit space $\mathcal M$. It can be done with the aid of the 
replacement of the variables $Q^A=F^A(Q^{\ast}(x),a)$ in which $a$  is constrained subsequently to $e$. 

In (\ref{4}), by ${\gamma}_{\mu \nu}(x)$   we  denote 
\[
d _{\mu \nu}(F(Q^{\ast}(x),e)=K^{A}_{\mu}(Q^{\ast}(x))\,G_{AB}(Q^{\ast}(x))K^B_{\nu}(Q^{\ast}(x)),
\]
where the $e$ is an identity element of the group $\mathcal G$.

The connection $A^{\mu}_i(x)$ is 
a pull-back of 
the Lie algebra-valued connection one-form 
$\Omega=\Omega ^{\alpha}\otimes e_{\alpha}$, which is given as follows:
\[
\Omega ^{\alpha}(Q)=d^{\alpha \beta}(Q)
G_{AB}(Q)K^B_{\beta }(Q)dQ^A.
\]
The matrix $\bar{u}^{\alpha}_{\beta}(a)$ is 
an inverse matrix to matrix
$\bar{v}^{\alpha}_{\beta}(a)=\frac{\partial {\Phi}^{\alpha}(b,a)}
{\partial b^{\beta}}\bigl|_{b=e}$.  $\Phi$ is the composition function 
of the group:
for  $c=ab$, $c^{\alpha}=~{\Phi}^{\alpha}(a,b)$.   

The determinant of the metric  $G_{AB}$ is equal to
\[
\det G_{AB}=\det h_{ij}(x)\,\det {\gamma }_{\alpha \beta
}(x)\,(\det \bar{u}_\rho ^\mu (a))^2.
\]

Performing the path integral transformation based on the transformation of the stochastic processes and on the nonlinear filtering stochastic differential equation,\footnote{This equation was used for the factorization of the path integral measure.} we  have obtained \cite{Storchak_1} the integral relation between the path integral given on the orbit space $\mathcal M$   and the path integral given on the total space  of the principal fiber bundle (the original manifold $\mathcal P$).  For the zero-momentum level reduction, 
this integral relation is 
\begin{equation}
{\gamma}(x_b)^{-1/4}{\gamma}(x_a)^{-1/4}
G_{M}(x_b,t_b;x_a,t_a)=\int_{\mathcal G}{G}_{P}
(\sigma (x_b)\theta ,t_b;\sigma(x_a),t_a)d\mu (\theta ),
\label{13}
\end{equation}
where ${\gamma}(x)=\det {\gamma }_{\alpha \beta}(x)$, $d\mu (\theta )$ is a normalized ($\int_{\mathcal G}d\mu (\theta )=1$)
invariant Haar measure on a group $\mathcal G$ and by 
$\sigma ^A(x)=f^A(x,e)$ we have denoted the local sections which allow us to express
the coordinates $Q^A$
in terms of $x^i$ and $\theta ^\alpha $: $Q^A=\sigma ^A(x)\theta $.

The Green function ${G}_{P}(Q_b,t_b;Q_a,t_a)$ represents the kernel of the evolution semigroup  (\ref{2}) which acts in the Hilbert space of functions with a scalar product 
$(\psi _1,\psi _2)=\int \psi _1(Q)\psi _2(Q)dv_{\mathrm P}(Q)$,
($dv_{\mathrm P}(Q)=\sqrt{G(Q)}dQ^1...dQ^{n_{\mathrm P}}$).

To obtain the probability representation of the kernel  ${G}_{P}$ one should to set  
$\varphi _0(Q)=G^{-1/2}(Q)\delta (Q-Q^{^{\prime }})$ in equation 
(\ref{2}).

The Green function $G_{M}$ is defined by the following path integral:
\begin{eqnarray}
 &&\!\!\!\!\!\!\!\!\!\!\!\!\!\!\!\!G_{M}(x_b,t_b;x_a,t_a)=\nonumber\\
&&\int d\mu ^{{x}}(\omega )\exp \left\{\frac 1{\mu
^2\kappa\, m}\int_{t_a}^{t_b}\tilde{V}({x}(u))du+\int_{t_a}^{t_b}J(
{x}(u))du\right\},
\label{6}
\end{eqnarray}
where $\tilde{V}(x)=V(f(x,a))$ (in case of the invariance of the potential term $V(Q)$) and the Jacobian of the reduction is 
\begin{equation}
J(x)=-\frac{\mu ^2\kappa }8\left[\triangle _{\mathrm {\cal M}}\ln {\gamma }+\frac
14h^{ni}\,
\frac {\partial \ln {\gamma }}{\partial x^n}
\frac {\partial \ln {\gamma }}{\partial x^i}\right].
\label{7}
\end{equation}
 
In (\ref{6}), the path integral measure $d\mu ^{{x}}$  is  related to the stochastic process $x_t$ which is given on the manifold 
$\mathcal M$. The local stochastic differential equations of the process $x_t$ are
\[
d{x}^i(t)=\frac 12\mu ^2\kappa \Bigl[\frac 1{\sqrt{h}}\frac \partial
{\partial x^n}(h^{ni}\sqrt{h})\Bigr]dt+\mu \sqrt{\kappa }X_{\bar{n}}^i(x(t))dw^{\bar{n}}(t).
\]
($h=\det h_{ij}(x)$,
$\sum^{n_{\mathrm {\cal M}}}_{{\bar{\mathrm n}}=1}{X}_{\bar{\mathrm n}}^{\mathrm i}
{X}_{\bar{\mathrm n}}^{\mathrm j}=h^{ij}$).

We note that the  semigroup determined by  
the kernel $G_M$ 
acts in the Hilbert space of functions with the following scalar product: 
$(\psi _1,\psi _2)=\int \psi _1(x)\psi _2(x)\,
dv_{\mathrm {\cal M}}(x)$ in which
$dv_{\mathrm {\cal M}}=\sqrt{h}\,dx^1...\, dx^{n_{\mathrm {\cal M}}}$.

The Hamilton operator $\hat H$ of the  Schr\"odinger equation can be obtained from the differential generator of this semigroup
by means of the relation 
$\hat H=-\frac{\hbar}{\kappa}{\hat H}_{\kappa}\bigl|_{\kappa =i}$.
It is equal to
\[
\hat{H}_{\kappa}=
\frac{\hbar \kappa}{2m}\triangle _{\mathrm {\cal M}}-\frac{\hbar \kappa}{8m}
\left[\triangle_{\mathrm {\cal M}}\ln {\gamma }+\frac 14(\nabla_{\mathrm {\cal M}}\ln {\gamma
})^2\right]+\frac{1}{\hbar \kappa}\tilde{V}.
\]

\section{The Jacobian}
Our present note will be concern with the geometrical representation  of the reduction Jacobian (\ref{7}).  
In order to get this representation 
we shall make use of a well-known formula 
\cite{Betounes,Cho} 
for the scalar curvature of the Riemannian manifold with the Kaluza--Klein metric:
\begin{eqnarray}
&&\!\!\!\!\!\!\!\!\!\!\!\!\!R_{\mathrm {\cal  P}}=R_{\mathrm {\cal M}}+R_{\mathrm {\cal G}}+\,\frac 14\, h^{ik}h^{mn}\,{\varphi}_{\alpha\beta}\,
{\tilde F}^{\alpha}_{im}{\tilde F}^{\beta}_{kn}\nonumber\\
&&\!\!\!\!\!\!\!\!\!\!\!\!\!
+\frac14\,h^{ij}{\varphi}^{\alpha\beta}{\varphi}^{\mu\nu}
\left[(\tilde{\mathcal D}_{i}{\varphi}_{\alpha\mu})(\tilde{\mathcal D}_{j}{\varphi}_{\beta\nu})+
(\tilde{\mathcal D}_{i}{\varphi}_{\alpha\beta})(\tilde{\mathcal D}_{j}{\varphi}_{\mu\nu})\right]+h^{ij}{\nabla}_{i}({\varphi}^{\alpha\beta}{\tilde{\mathcal D}}_{j}{\varphi}_{\alpha\beta}),
\label{R}
\end{eqnarray}
where $R_{\mathrm {\cal M}}$ is a scalar curvature of the orbit space $\cal M$, 
$R_{\mathrm {\cal G}}$ is a scalar curvature of the orbit ${\cal G}_x$ endowed with the induced metric. 

This formula can be  obtained if for the  calculation of the Christoffel coefficients 
one takes  a special basis, the horisontal lift basis,  which turns the Kaluza--Klein metric into the block diagonal form. In our notation this basis is formed by $({\tilde H}_i, L_{\mu})$, where ${\tilde H}_i={\partial}_i-{\tilde A}^{\mu}_i(x,a)L_{\mu}$, ${\tilde A}^{\mu}_i(x,a)={\bar {\rho}}^{\mu}_{\nu}(a) A^{\nu}_i(x)$, in which  
${\bar{\rho}}^{\mu}_{\nu}$ is an inverse matrix to 
the matrix 
${\rho}^{\mu}_{\nu}={\bar u}^{\mu}_{\sigma}\,v^{\sigma}_{\nu}$  of an adjoint representation of the group $\cal G$.
 $L_{\mu}=v^{\sigma}_{\mu}(a){\partial}/{\partial} a^{\sigma}$
($[L_{\alpha},L_{\beta}]=c^{\mu}_{\alpha \beta}L_{\mu}$).
In this basis, ${\tilde G}({\tilde H}_i,{\tilde H}_j)=h_{ij}(x)$ and 
${\tilde G}(L_{\alpha},L_{\beta})={\varphi}_{\alpha \beta}(x,a)={\rho}^{\mu}_{\alpha}(a){\rho}^{\nu}_{\beta}(a){\gamma}_{\mu\nu}(x)$.

The Riemannian curvature and the scalar curvature are defined by the following formulas. The Riemann curvature operator $\Omega$ for the connection $\nabla$  is given by
\[
 {\Omega}(X,Y)=\left[{\nabla}_X,{\nabla}_Y\right]
-{\nabla}_{\left[X,Y\right]},
\]
 and the Riemann tensor is $R(X,Y,Z,Z^{'})=G({\Omega}(X,Y)Z,Z^{'})$.
Contracting in  repeated indices in the Riemann tensor, one can obtain the Ricci tensor
$R_{AC}=R_{AMC}{}^M$ and then the scalar curvature (\ref{R}) in which 
 $\tilde{\mathcal D}_i$ is a covariant derivative, ${\nabla}_i$ is the gauge and general covariant derivative and ${\tilde F}^{\alpha}_{im}(x,a)={\partial}_i {\tilde A}^{\alpha}_m-{\partial}_m {\tilde A}^{\alpha}_i+c^{\alpha}_{\mu\nu}\, {\tilde A}^{\mu}_i {\tilde A}^{\nu}_m$. 

In our basis $(\frac{\partial}{\partial x^i},\frac{\partial }{\partial a^{\mu}})$, (\ref{R}) can be rewritten as follows:\footnote{We have used the condition $c^{\kappa}_{\mu\kappa}=0$ which is valid for the compact semisimple group.}
\begin{eqnarray}
R_{\mathrm {\cal P}}&=&R_{\mathrm {\cal M}}+R_{\mathrm {\cal G}}+\frac 14\,{\gamma}_{\mu \nu}\,
F^{\mu}_{\,ij}\,F^{\nu\,ij}+\frac14\,h^{ij}\,{\gamma}^{\mu\sigma}{\gamma}^{\nu\kappa}
\left({\mathcal D}_{i}{\gamma}_{\mu\nu}\right)\left({\mathcal D}_{j}{\gamma}_{\sigma\kappa}\right)\nonumber\\
&&+\,
\frac14\,h^{ij}\left({\gamma}^{\mu\nu}
{\partial}_{i}{\gamma}_{\mu\nu}\right)\left(
{\gamma}^{\sigma\kappa}{\partial}_{j}{\gamma}_{\sigma\kappa}\right)
-h^{ij}\,{\gamma}^{\mu\sigma}{\gamma}^{\nu\kappa}
\left({\partial}_{i}{\gamma}_{\mu\nu}\right)\left({\partial}_{j}{\gamma}_{\sigma\kappa}\right)
\nonumber\\
&&+\,h^{ij}\,{\gamma}^{\mu\nu}
\left({\partial}_{i}\,{\partial}_{j}{\gamma}_{\mu\nu}-{\mathrm{\Gamma}}^k_{ij}\,{\partial}_{k}{\gamma}_{\mu\nu}\right),
\label{8}
\end{eqnarray}
where the scalar curvature of the orbit  $R_{\mathrm {\cal G}}=\frac12{\gamma}^{\mu\nu} c^{\sigma}_{\mu \alpha} c^{\alpha}_{\nu\sigma}+
\frac14 {\gamma}_{\mu\sigma}{\gamma}^{\alpha\beta}{\gamma}^{\epsilon\nu}
c^{\mu}_{\epsilon \alpha}c^{\sigma}_{\nu \beta}$,
$F^{\alpha}_{ij}(x)$ is related to ${\tilde F}^{\alpha}_{ij}$ by ${\tilde F}^{\alpha}_{ij}(x,a)={\bar{\rho}}^{\alpha}_{\mu}(a)
F^{\mu}_{ij}(x)$ and
the covariant derivative ${\mathcal D}_i{\gamma}_{\mu\nu}$ is given by
$
 {\mathcal D}_{i}{\gamma}_{\mu\nu}={\partial}_i{\gamma}_{\mu\nu}-
c^{\kappa}_{\sigma \mu}A^{\sigma}_i{\gamma}_{\kappa\nu}-
c^{\kappa}_{\sigma \nu}A^{\sigma}_i{\gamma}_{\mu\kappa}.
$

Comparing  the second and third line of (\ref{8}) with 
the expressions $\tilde J$ standing  
under the square bracket in (\ref{7}), we can see that they are equal.  It allows us to represent 
the term $\tilde J=\left[\triangle _{\mathrm M}\ln {\gamma }+\frac
14h^{ni}\,
({\partial}_n \ln {\gamma })
({\partial}_i \ln {\gamma })\right]$ in the following form: 
\begin{equation}
  \tilde J=R_{\mathrm {\cal P}}-R_{\mathrm {\cal M}}-R_{\mathrm {\cal G}}-\frac 14\,{\gamma}_{\mu \nu}\,
F^{\mu}_{\,ij}\,F^{\nu\,ij}-\frac14\,h^{ij}\,{\gamma}^{\mu\sigma}{\gamma}^{\nu\kappa}
\left({\mathcal D}_{i}{\gamma}_{\mu\nu}\right)\left({\mathcal D}_{j}{\gamma}_{\sigma\kappa}\right).
\label{9}
\end{equation}

To obtain the geometrical representation for the  last term in (\ref{9}), we make use of the second fundamental form of the orbit.
In the total space of the fibre bundle this form is determined as 
\[
 j^C_{\alpha\beta}(Q)={\mathrm {\Pi}}^C_E(Q) \left({\nabla}_{K_{\alpha}}K_{\beta}\right)^E(Q).
\]
Projecting the second fundamental form, taken at $(x^i,a^{\alpha})$, onto the direction which is parallel to the orbit space, we get
\begin{eqnarray}
&&{\tilde G}^{in}\,{\tilde G}\left(j^C_{\alpha\beta}(Q)\frac{\partial}{\partial Q^C},
\frac{\partial}{\partial x^i}\right)\frac{\partial}{\partial x^n}=
\frac12\,{\rho}^{{\alpha}'}_{\alpha}(a){\rho}^{{\beta}'}_{\beta}(a)
\left({\nabla}_{K_{{\alpha}'}}K_{{\beta}'}+
{\nabla}_{K_{{\beta}'}}K_{{\alpha}'}\right)^E
\nonumber\\
&&\;\;\;\;\;\times\;{}^HG_{EB}(Q^{\ast}(x))\,{Q^{\ast}}^B_m(x)\,h^{mn}(x)\frac{\partial}{\partial x^n},
\label{10}
\end{eqnarray}
where ${\tilde G}$ is the Kaluza--Klein metric (\ref{4}) 
  and the terms in the bracket on the right--hand side of the obtained equality depend on $Q^{\ast}(x)$.

Changing the coordinates $Q^A$ for the coordinates $(x^i, a^{\mu})$ 
in the identity
\[
\frac{\partial \,d_{\alpha \beta}(Q)}{\partial Q^C}=-G_{CE}(Q)
\left({\nabla}_{K_{{\alpha}}}K_{{\beta}}+
{\nabla}_{K_{{\beta}}}K_{{\alpha}}\right)^E(Q),\]
we obtain that
\[
{\mathcal D}_{i}{\gamma}_{\alpha\beta}(x)=- 
\,{}^HG_{CE}(Q^{\ast}(x))\,{Q^{\ast}}^C_i(x)
\left({\nabla}_{K_{{\alpha}}}K_{{\beta}}+
{\nabla}_{K_{{\beta}}}K_{{\alpha}}\right)^E(Q^{\ast}(x)).
\]
Therefore,  the second fundamental form of the orbit,
 restricted to the orbit space, 
 is equal to
\[
j^n_{\alpha\beta}(x)=-\frac12\, h^{ni}(x)\,{\mathcal D}_{i}{\gamma}_{\alpha\beta}(x),
\]
and we come to the following representation of $\tilde J(x)$:
 \begin{equation}
  \tilde J=R_{\mathrm {\cal P}}-R_{\mathrm {\cal M}}-R_{\mathrm {\cal G}}-\frac 14\,{\gamma}_{\mu \nu}\,
F^{\mu}_{\,ij}\,F^{\nu\,ij}-\,h_{kn}\,{\gamma}^{\alpha
\mu}\,{\gamma}^{\beta\nu}\,j^k_{\alpha\beta}\,j^n_{\mu\nu}.
\label{11}
\end{equation}
The Hamilton operator of the  Scr\"odinger equation 
on the reduced manifold $\mathcal M$
will be
\[
\hat{H}=
-\frac{{\hbar}^2 }{2m}\triangle _{\mathrm {\cal M}}+\frac{{\hbar}^2 }{8m}
\left[R_{\mathrm {\cal P}}-R_{\mathrm {\cal M}}-R_{\mathrm {\cal G}}-\frac 14\,{\gamma}_{\mu \nu}\,
F^{\mu}_{\,ij}\,F^{\nu\,ij}-\, 
||j||^2
\right]+\tilde{V}.
\]
A similar formula for the quantum corrections to the Hamiltonian was obtained by Gawedzki in \cite{Gawedzki}.

In  case of the reduction onto the non-zero momentum level  ($\lambda \neq 0$) \cite{Storchak_1}  we obtain the following representation for the Hamilton operator:
\[
{\hat{H}}^{\cal E}=
-\frac{{\hbar}^2 }{2m}
\left({\triangle}^{{\mathcal E}}\right)^{\lambda}_{pq}
+{\gamma}^{\alpha\nu} (J_{\alpha})^{\lambda}_{pq^{'}}(J_{\nu})^{\lambda}_{q^{'}q}
+\frac{{\hbar}^2 }{8m}\,[\,{\tilde J}(x)\,]\, {\rm I}^{\lambda}_{pq}
+\tilde{V}\,{\rm I}^{\lambda}_{pq},
\]
where $(J_{\alpha})^{\lambda}_{pq}$ are infinitesimal generators of the irreducible representation $T^{\lambda}$, 
${\rm I}^{\lambda}_{pq}$ is a unity matrix.
The horizontal Laplacian ${\triangle}^{{\mathcal E}}$ is given  by the formula:
\begin{eqnarray*}
\left({\triangle}^{{\mathcal E}}\right)^{\lambda}_{pq}
&=&{\sum}_{\bar k=1}^{n_{\cal M}}
\left({\nabla}^{\mathcal E}_{X^i_{\bar k}{\rm e_i}}
{\nabla}^{\mathcal E}_{X^j_{\bar k}{\rm e_j}}-
{\nabla}^{\mathcal E}_{{\nabla}^{\mathcal M}_{X^i_{\bar k}{\rm e_i}} {X^j_{\bar k}{\rm e_j}}}\right)^{\!\!\lambda}_{\!\!pq}\nonumber\\
&=&{\triangle}_{\cal M}\,
{\rm I}^{\lambda}_{pq} 
+2h^{ij}({\rm {\Gamma}^{\mathcal E}})^{\lambda}_{ipq}\,{\partial}_j\nonumber\\
&&-\,h^{ij}\left[{\partial}_i ({\rm {\Gamma}^{\mathcal E}})^{\lambda}_{jpq}
-({\rm {\Gamma}^{\mathcal E}})^{\lambda}_{ip{q}^{'}}({\rm {\Gamma}^{\mathcal E}})^{\lambda}_{j{q}^{'}q}+({\rm {\Gamma}^{\cal M}})^{m}_{ij}\,
({\rm{\Gamma}}^{\mathcal E})^{\lambda}_{mpq}\right],
\end{eqnarray*}
in which $X^i_{\bar k}$ is defined by the local equality  ${\sum}_{\bar k=1}^{n_{\cal M}}X^i_{\bar k}X^j_{\bar k}=h^{ij}$
  and 
$({\rm {\Gamma}^{\mathcal E}})^{\lambda}_{ipq}=A^{\alpha}_n (J_{\alpha})^{\lambda}_{pq}$ are  connection coefficients  of the associated bundle ${\cal E}=P{\times}_{\cal G} V_{\lambda}$.
The operator ${\hat{H}}^{\cal E}$ acts in the space of the section of this bundle with the scalar product 
\[
 ({\psi}_1,{\psi}_2)={\int}_{\!\!\!\!\cal M}\langle {\psi}_1,{\psi}_2{\rangle}_{V_{\lambda}}dv_{\cal M}(x),
\]
${\langle}\cdot ,\cdot{\rangle}_{V_{\lambda}}$ is an internal scalar product.

In conclusion it should be noted that the obtained representation of  the Jacobian (\ref{11}) depends on the definition of the curvature tensor. Using an another definition, such as, for example, in \cite{Misner,DeWitt,Chiang}, one  
may get the following relation:
\[
R^{'}_{\mathrm {\cal P}}=R^{'}_{\mathrm {\cal M}}+R^{'}_{\mathrm {\cal G}}-\frac 14\,{\gamma}_{\mu \nu}\,
F^{\mu}_{\,ij}\,F^{\nu\,ij}-\,||j||^2 -\tilde J.
\]

\end{document}